\documentclass[prb,twocolumn,preprintnumbers,amsmath,amssymb,epsfig,superscriptaddress]{revtex4}

\usepackage{epsfig}
\usepackage{mathrsfs}
\usepackage{amssymb}

\usepackage{graphicx}
\usepackage{dcolumn}
\usepackage{bm}

\begin{document}

\title{Josephson junction transmission lines as tunable artificial crystals}

\author{Carsten Hutter}
\affiliation{Department of Physics, Stockholm University, AlbaNova
University Center, SE--106 91 Stockholm, Sweden}
\affiliation{Nanostructure Physics, Royal Institute of Technology,
SE--106 91 Stockholm, Sweden}

\author{Erik A. Thol{\'e}n}
\affiliation{Nanostructure Physics, Royal Institute of Technology,
SE--106 91 Stockholm, Sweden}

\author{Kai Stannigel}
\email[Present address: Institute for Theoretical Physics, University of Innsbruck, and Institute for Quantum Optics and Quantum Information, 6020 Innsbruck, Austria.]{}
\affiliation{Department of Physics, Stockholm University, AlbaNova
University Center, SE--106 91 Stockholm, Sweden}

\author{Jack Lidmar}
\affiliation{Theoretical Physics, Royal Institute of Technology,
SE--106 91 Stockholm, Sweden}

\author{David B. Haviland}
\affiliation{Nanostructure Physics, Royal Institute of Technology,
SE--106 91 Stockholm, Sweden}

\date{\today}

\begin{abstract}
We investigate one-dimensional Josephson junction arrays with generalized unit cells as a circuit approach to engineer microwave band gaps. An array described by a lattice with a basis can be designed to have a gap in the electromagnetic spectrum, in full analogy to electronic band gaps in diatomic or many-atomic crystals.  We derive the dependence of this gap on the array parameters in the linear regime, and suggest experimentally feasible designs to bring the gap below the single junction plasma frequency. The gap can be tuned in a wide frequency range by applying external flux, and it persists in the presence of small imperfections.
\end{abstract}

\maketitle

\section{Introduction}

The design of Josephson junction circuits in an appropriate electromagnetic environment\cite{physReports1990schoen} is currently of great interest in the context of circuit QED and qubit design \cite{rmp2001makhlin, pra2008zhou,prb2008rakhmanov}.  The quantum mechanical nature of these electronic circuits is often described by analogy, where individual circuit elements can be thought of as ``artificial atoms'' whose intrinsic properties can be designed by the quantum circuit engineer.  A natural extension of this analogy is to view periodic arrays of circuit elements as fully designable, tunable artificial crystals\cite{pra2008zhou,THz,prb2008rakhmanov}, or engineered metamaterials \cite{pra2008zhou,ricci,prb2006du,apl2007lazarides,THz,prb2008rakhmanov}.  Metamaterials based on optical plasma resonances in metallic nanostructures are presently of great interest\cite{science2006ozbay}, but their microwave counterparts are perhaps even more interesting when superconductors are used to realize the metamaterial, due to the absence of dissipation for frequencies below the superconducting energy gap \cite{ricci,prb2006du,apl2007lazarides,THz}.  

Both the classical and the quantum electrodynamics of these periodic structures is extremely rich when Josephson tunnel junctions are used to build the metamaterial.  Periodic arrays of Josephson junctions have specifically been the subject of numerous studies as a model system for quantum phase transitions (for a review see Ref.  \onlinecite{physReports2001fazio}).  While much of the early work in this field concerns 2d-JJAs, the quantum behavior of 1d-JJAs has also been investigated \cite{prb1984bradley, rmp1997sondhi, prl2002matveev, naturePhysics2010pop}, and  1d-JJAs have also been described in the context of quantum metamaterials built from integrated qubit chains \cite{prb2008rakhmanov} and the transfer of quantum information with on-chip transmission lines~\cite{prb2005romito}.   Other studies treat 1d-JJAs classically, where the nonlinear Josephson inductance is used to amplify signals at the quantum limit~\cite{apl1996yurke,naturePhysics2008castellanos-beltran}.  The large linear inductance of the 1d-JJA has recently been used to realize a charge qubit immune to low-frequency charge noise \cite{prl2009koch, science2009manucharyan}.  The classical phase dynamics of regular 1d-JJAs was also studied for the development of the 10 Volt Josephson voltage standard, where the focus was on the nonlinear dynamics of a driven array, in order to understand the boundary between periodic and chaotic response \cite{progPhysics1996kautz, physlett2002dhamala}.

There exists however a gap in the literature concerning the classical electrodymamics of 1d-JJAs, which we address in this article.   Here we examine the simple linear electrodynamics of a 1d-JJA when the array is described by a {\em lattice with a basis}.  The presence of a basis in the one-dimensional lattice causes the appearance of a gap in the electromagnetic spectrum of the array, in analogy to a many-atomic crystal. Our interest is to simulate experimentally realizable designs where such more complex unit cells are used to control the dispersion relation and band gap. Specifically, designs consisting of a basis with two different junctions (or SQUIDs) and different capacitances to ground are considered.  By using SQUIDs in the unit cell, the gap can be tuned in a wide range using an external magnetic field.  We derive the dispersion relation of an infinite array with unit cells having two different junctions in the basis, as depicted in Fig. 1. 

\begin{figure}
\begin{centering}%
\hspace{-0.0cm}
\begin{minipage}[b][1\totalheight][c]{1\columnwidth}%
\begin{minipage}[c][1\totalheight]{0.02\columnwidth}%
  (a)%
\end{minipage}%
\hspace{-0.1cm}%
\begin{minipage}[t][1\totalheight][c]{0.8\columnwidth}%
\vspace{-0.2cm}
 \begin{center}\vspace{0cm}\includegraphics[scale=0.25]{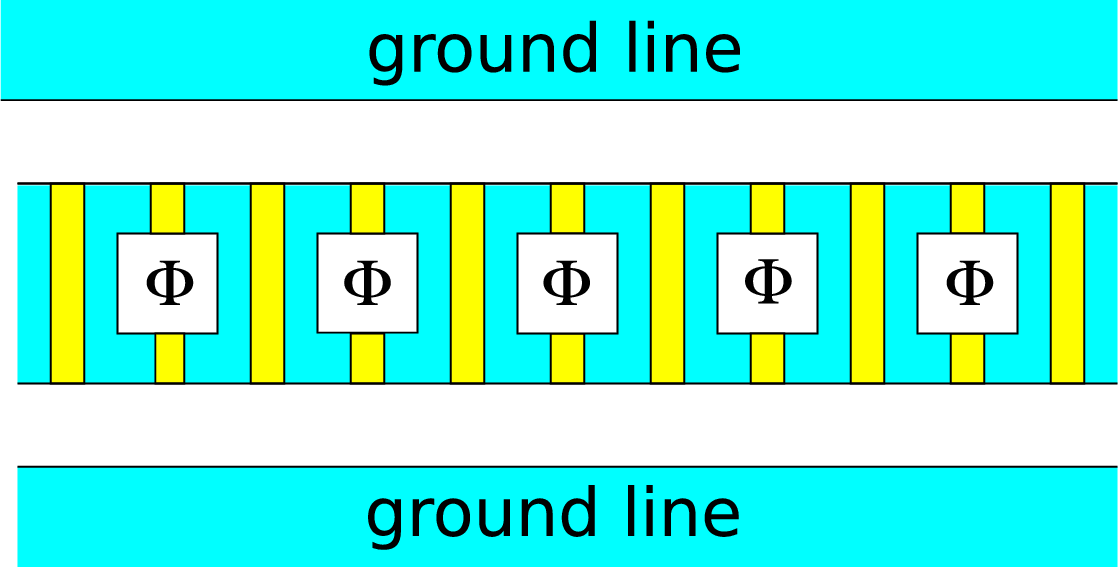}\par\end{center}%
\end{minipage}\\%
\vspace{0.2cm}
\begin{minipage}[c][1\totalheight]{0.02\columnwidth}%
  (b)%
\end{minipage}%
\hspace{-0.1cm}%
\begin{minipage}[t][1\totalheight][c]{0.43\columnwidth}%
\vspace{-0.2cm}
  \begin{center}\vspace{0cm}\includegraphics[scale=0.31]{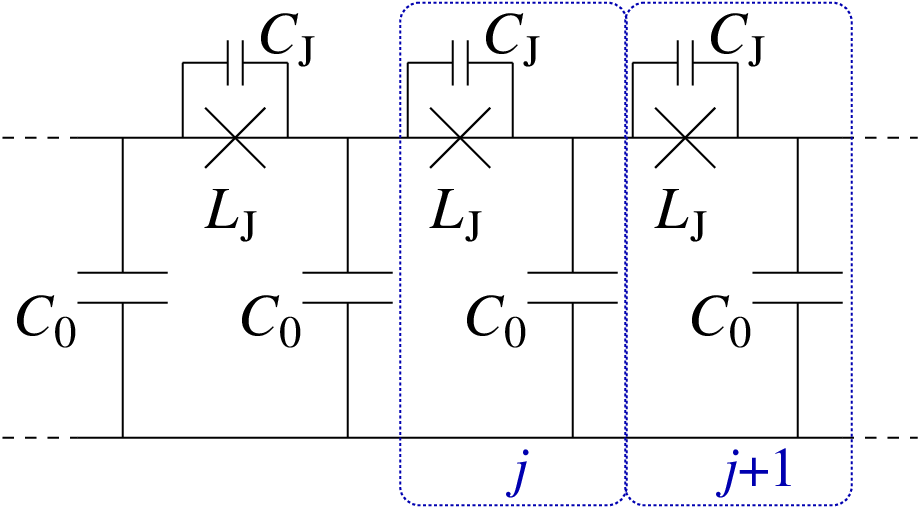}\par\end{center}%
\end{minipage}%
\hspace{1.4cm}%
\begin{minipage}[t][1\totalheight]{0.02\columnwidth}%
  (c)%
\end{minipage}%
\hspace{0.1cm}%
\begin{minipage}[t][1\totalheight][c]{0.36\columnwidth}%
\vspace{-0.2cm} 
 \begin{center}\vspace{0.02cm}\includegraphics[scale=0.31]{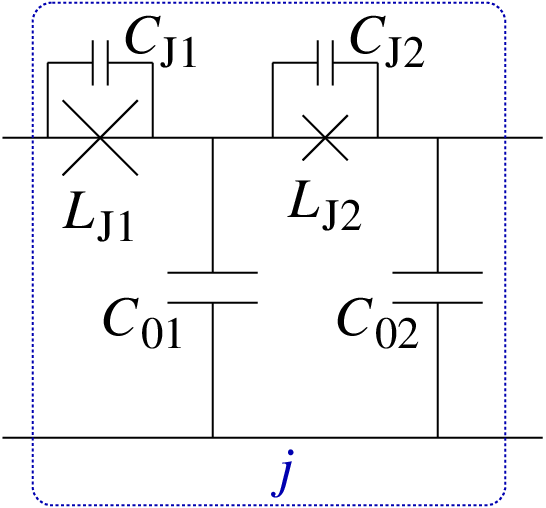}\par\end{center}%
\end{minipage}%
\end{minipage}%
\par\end{centering}
\caption{
\label{fig:twoModels}
(Color online)
(a)  A sketch of a coplanar transmission line where the center conductor is a regular 1d-JJA with a basis containing both a single junctions and a SQUID loops.  (b) A lumped element circuit model of a regular 1d-JJA, where each unit cell $j$ consists of a Josephson junction with its parallel capacitance and a capacitance to ground.  (c)  A generalized model where each unit cell $j$ consists of two different junctions and capacitances to ground, suitable to study the design in Fig.~1a.
}
\end{figure}

This Josephson junction analog of the diatomic chain \cite{book1976ashcroft} has more parameters than its atomic counterpart,  allowing for control in the design of the dispersion relation.  As with the diatomic chain, the dispersion relation has a band gap as shown in Fig.~\ref{fig:dispersionRelationTwoJunctions}, i.e., a frequency region in which no propagating modes appear and the real part of the impedance vanishes.  The gap appears at frequencies of the order of the plasma frequencies of the individual junctions. We investigate the parameter dependence of the band gap and possibilities to lower it to the experimentally accessible frequency region, appropriate for integration with qubit designs. Extending this approach, one sees that an even wider parameter space can be achieved with unit cells having more than two junctions or SQUIDs, resulting in more branches in the linear dispersion relation.  While we focus on the case with two different junctions per unit cell, we give in the appendix a more general treatment which allows arbitrary, non-identical unit cells in the linear regime.  We use this general approach to show that the gap persists in finite, short arrays with two-junction unit cells in the presence of a small parameter spread ($5\%$ standard deviation), and we present simulations for a transmission experiment with realistic boundary conditions.

The linear approximation restricts the applicability of our model to junctions where quantum tunneling of the phase can be neglected, which is realized when the Josephson energy dominates over the charging energy. Such junctions  have a comparatively large area and therefore have the advantage that they can be fabricated with a low relative spread of parameters.  Junctions in the phase regime are approximately described by their linear behavior if the current flowing in the junctions is much less than the critical current. The array can then be regarded as a complex transmission line  with a non-trivial, gapped dispersion relation. A resonator made from a finite-length transmission line with such an array could find use in circuit cavity QED\cite{nature2004wallraff,nature2007schuster} for strongly coupling to the Josephson plasma modes.  Nonlinear corrections, briefly discussed in the Appendix, can be used to realize parametric amplification~\cite{prl1988yurke,apl1996yurke,apl2007castellanos-beltran,apl2007tholen,naturePhysics2008castellanos-beltran,jlt2006yurke} and quantum noise squeezing~\cite{prl1990movshovich,naturePhysics2008castellanos-beltran,jlt2006yurke}.

\section{Josephson junction arrays with two-junction unit cells}
In Fig. 1a we show a regular 1d-JJA in a coplanar transmission line geometry, with unit cells consisting of one simple junction in series with a SQUID. Each SQUID consists of two parallel junctions with Josephson energy $E_{\rm J0}$.  When pierced by a flux $\Phi=BA_{\rm S}$, where $B$ is an applied magnetic field, and $A_{\rm S}$ is the effective area of the SQUID loop, each SQUID is effectively identical to a single junction with tunable Josephson energy $E_{\rm J}=2E_{\rm J0}\cos(2\pi|\Phi|/\Phi_0)$, and we can thus regard the design in Fig.~1a as a design with two different junctions per unit cell. Linearizing the Josephson relation, each effective junction is described by its capacitance $C_{\rm J}$ and the linear Josephson inductance $L_{\rm J}=\Phi_0^2/(4\pi^2 E_{\rm J})$ with superconducting flux quantum $\Phi_0=h/2e$.  We consider situations were quasiparticle tunneling can be neglected, with the voltage drop across each junction less than the superconducting energy gap, $V<2\Delta / e$.  We also introduce the plasma frequency  $\omega_{\rm p}=1/\sqrt{L_{\rm J}C_{\rm J}}$ for use later on.

In Fig.~1b we show a simple model of a 1d-JJA with identical junctions, taking into account a capacitance to ground $C_0$.  Note that in the limit $C_{\rm J}\rightarrow 0$ the model reduces to the discrete, lumped element model of a transmission line for transverse electromagnetic waves.  In Fig.~1c we show the generalization studied in this article, where the array consists of a lattice of unit cells each consisting of a basis of two Josephson junctions, for which we introduce an additional index $1$ or $2$ to the parameters above, see Fig.~1c.

\begin{figure}
\begin{centering}%
\hspace{-0.95cm}
\begin{minipage}[b][1\totalheight][c]{0.99\columnwidth}%
\begin{minipage}[c][1\totalheight]{0.02\columnwidth}%
  (a)%
\end{minipage}%
\hspace{-0.15cm}%
\begin{minipage}[t][1\totalheight][c]{0.43\columnwidth}%
  \begin{center}
  \vspace{-0.15cm}\includegraphics[scale=0.5]
  {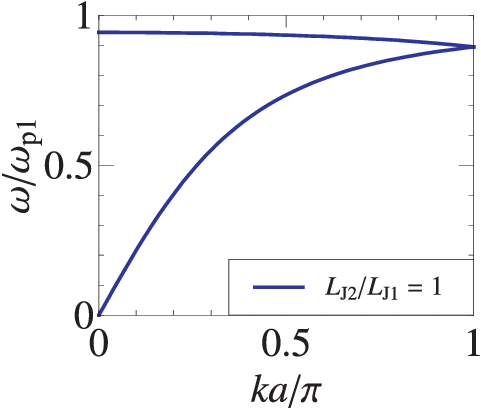}
  \par\end{center}%
\end{minipage}%
\hspace{0.70cm}%
\begin{minipage}[t][1\totalheight]{0.02\columnwidth}%
  (b)%
\end{minipage}%
\hspace{-0.15cm}%
\begin{minipage}[t][1\totalheight][c]{0.41\columnwidth}%
  \begin{center}\vspace{-0.15cm}\includegraphics[scale=0.5]
  {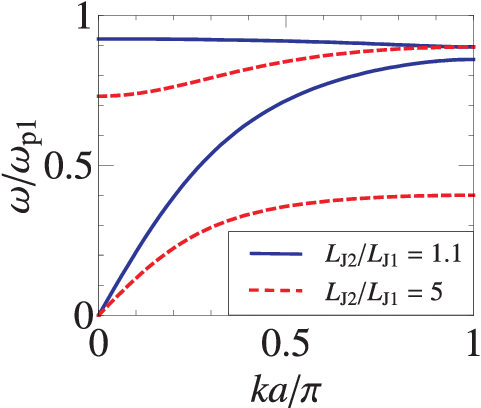}
  \par\end{center}%
\end{minipage}%
\end{minipage}%
\par\end{centering}
\caption{ \label{fig:dispersionRelationTwoJunctions} (Color online)
Dispersion relation for the model with two junctions when the
junction parameters are (a) symmetric, $L_{\rm J2}/L_{\rm J1}=1$,
and (b) asymmetric, with weak ($L_{\rm J2}/L_{\rm J1}=1.1$) and
stronger ($L_{\rm J2}/L_{\rm J1}=5$) asymmetry. In all cases we used
$C_{\rm J1}=C_{\rm J2}$ and $C_{01}/C_{\rm J1}=C_{02}/C_{\rm
J2}=0.5.$}
\end{figure}

For non-identical junctions in the unit cell one expects a band gap in the dispersion relation of the transmission line, in analogy to a diatomic chain.  This gap is shown in Fig.~\ref{fig:dispersionRelationTwoJunctions} for two different values of asymmetry parameter $L_{J2}/L_{J1}$.  The gap does not appear in systems with simple unit cells as in Fig.~1b, where each unit cell has only one independent degree of freedom due to loop constraints. In this case, linearization of the equations of motion in small value of phase difference across each junction approximates the system as coupled harmonic oscillators and a traveling wave ansatz yields a single branch in the dispersion relation.  This branch will have an upper cutoff frequency due to the discreteness of the model, but no second branch and no gap.  However, if we consider different unit cells consisting of two original cells, each of the new cells has in general two independent degrees of freedom. This results in a representation of the dispersion relation where the original branch is mirrored at half the Brillouin zone, and thus appears as two branches as shown in Fig.~\ref{fig:dispersionRelationTwoJunctions}a. Here, we used the length $a$ for the new unit cell. If an asymmetry of parameters is introduced within each unit cell, a splitting in the dispersion relation into ``acoustic'' and ``optical'' bands occurs as shown in Fig.~\ref{fig:dispersionRelationTwoJunctions}b.

As shown in Appendix A, the dispersion relation for an infinite array in the linear approximation is
\begin{equation}
\label{eq:dispTwo}
 [\omega_\pm(k)]^2=\frac{B}{2A} \pm
\frac{\sqrt{B^2-4AC}}{2A}\ ,
\end{equation}
where
\begin{eqnarray}
\label{eq:ABC}
A&=&(C_{01}+C_{02})(C_{\rm J1}+C_{\rm J2})+C_{01}C_{02}+C_{\rm J1}C_{\rm J2}\beta_k\nonumber\\
B&=&\frac{C_{01}+C_{02}}{L_{12}} +(\frac{C_{\rm J1}}{L_{\rm
J2}}+\frac{C_{\rm J2}}{L_{\rm J1}})\beta_k
\nonumber\\
C&=&\frac{\beta_k}{L_{\rm J1} L_{\rm J2}}\ ,
\end{eqnarray}
with $1/L_{12}=1/L_{\rm J1}+1/L_{\rm J2}$ and $\beta_k=2[1-\cos(ka)]$. Since $A>0$, one sees from the defining Eq.~(\ref{eq:dispTwo}) that $\omega_+(k)\ge \omega_-(k)$ for any wave vector $k$.

The lower and upper edge of the gap are defined as
\begin{eqnarray}
\omega_{\rm gL}&=&\max_k\{\omega_-(k)\}\, ,\nonumber\\
\omega_{\rm gU}&=&\min_k\{\omega_+(k)\}\, ,
\end{eqnarray}
respectively. In the following we consider positive inductances and
capacitances, for which the maximum of the lower band edge always
appears at wave vectors $ka=\pi$, where $\beta_k=4$, while
the upper band edge can appear at $ka=0$ 
or $ka=\pi$, depending on the parameters, cf. Fig.~\ref{fig:dispersionRelationTwoJunctions}b. 
More
explicitly, we find
\begin{eqnarray}
\label{eq:extrema}
\omega_{\rm gL}&=&\omega_-(k=\pi/a)\ ,\nonumber\\
\omega_{\rm gU}&=&\left\{
\begin{array}{lll}
\omega_+(k=0)=\frac{1}{\sqrt{L_{12}C_\Sigma}}&\mbox{ for }&\xi_1\cdot\xi_2\le 0\nonumber\\
\omega_+(k=\pi/a)&\mbox{ for }&\xi_1\cdot\xi_2\ge 0\ ,
\end{array}
\right.
\end{eqnarray}
where we defined $C_\Sigma=C_{\rm J1}+C_{\rm
J2}+C_{01}C_{02}/(C_{01}+C_{02})$, and
\begin{eqnarray}
\xi_1&=&\left[C_{02}C_{{\rm J}2}+C_{01}(C_{02}+C_{{\rm
J}2})\right]L_{{\rm J}2}\nonumber\\&&-(C_{01}+C_{02})C_{{\rm
J}1}L_{{\rm
J}1}\nonumber\\
\xi_2&=&\left[C_{02}C_{{\rm J}1}+C_{01}(C_{02}+C_{{\rm
J}1})\right]L_{{\rm J}1}\nonumber\\&&-(C_{01}+C_{02})C_{{\rm
J}2}L_{{\rm J}2}\ .
\end{eqnarray}
\begin{figure}
\hspace{0.8cm}
\begin{centering}%
\hspace{-0.0cm}
\begin{minipage}[b][1\totalheight][c]{0.97\columnwidth}%
  \begin{center}\vspace{0cm}\hspace{0.2cm}\includegraphics[scale=0.5]{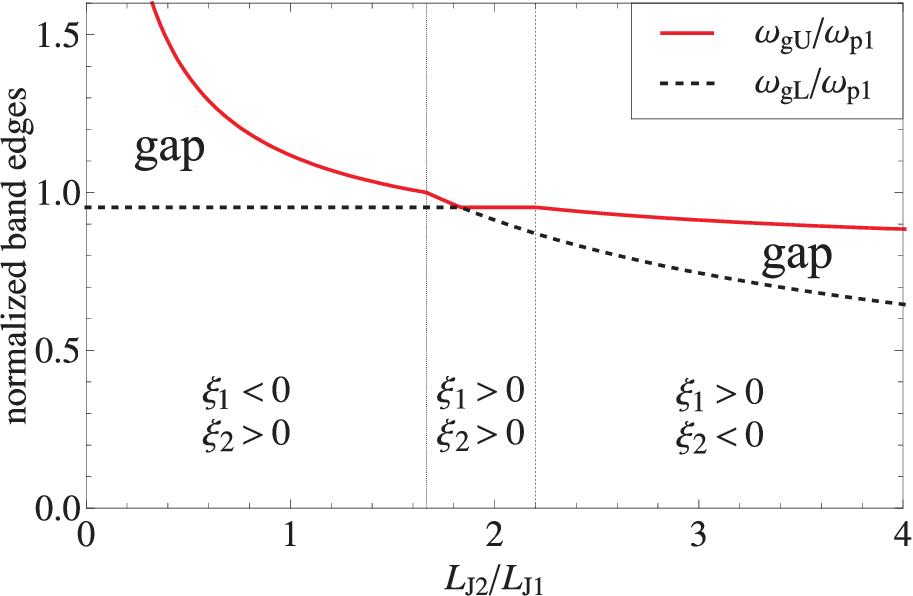}\par\end{center}%
\end{minipage}%
\par\end{centering}
\caption{ \label{fig:dispersionLowEnergy} (Color online) The upper
and lower band edge in dependence of the ratio $L_{\rm J2}/L_{\rm
J1}$. We used $C_{\rm J2}/C_{\rm J1}=0.5$ and
$C_{01}=C_{02}=0.2C_{\rm J1}$. Thin vertical lines mark where
$\xi_1$ or $\xi_2$ change sign.}
\end{figure}

We show in Fig.~\ref{fig:dispersionLowEnergy} how the gap can be
moved in frequency if one can control the ratio of effective
Josephson inductances $L_{\rm J2}/L_{\rm J1}$. The gap vanishes,
$\omega_{\rm gL}=\omega_{\rm gU}$, if both $C_{01}=C_{02}\equiv C_0$
and
\begin{equation}
\label{eq:noGap}
L_{\rm J1}C_{\rm J1}=L_{\rm J2}C_{\rm
J2}+C_0(L_{\rm J2}-L_{\rm J1})/2\ .
\end{equation}
Kinks in the plot appear at the points, where either $\xi_1$ or
$\xi_2$ change sign. According to Eq.~(\ref{eq:extrema}) this
corresponds to switching the position of the minima of $\omega_+(k)$
between $k=0$ and $k=\pi/a$, which can be shown to be realized by a
flat (constant in $k$) upper branch $\omega_+(k)$ at these points.

Tunable inductances as in Fig.~\ref{fig:dispersionLowEnergy} can be
achieved by employing one SQUID (Fig. 1a) or two SQUIDs in each unit cell. A design
with one junction and one SQUID per unit cell, as shown in Fig. 1a, has the advantage that
one of the two Josephson energies, which are inversely proportional to the respective inductances, can be tuned continuously without
changing the other Josephson energy. A design with two SQUIDs per unit
cell has different advantages. Clearly, one can then tune both
Josephson energies. If one chooses different areas $A_{\rm S1}$ and
$A_{\rm S2}$  for the two SQUIDs, a change in magnetic field $\Delta
B=\Phi_0/A_{\rm S1}$ leaves the Josephson energy of the first SQUID
invariant, while it changes that of the second SQUID. In this sense,
one can tune both Josephson energies independently with only one
common magnetic field\cite{prl2006corlevi}. Thus, this design is preferable in
experiments which test many combinations of Josephson inductances $(L_{\rm
J1}, L_{\rm J2})$, while the aforementioned design is better if one
needs a continuous change of one inductance, which  might become
important for applications.

\begin{figure}[b]
\begin{centering}%
\hspace{-0.5cm}
\begin{minipage}[b][1\totalheight][c]{0.99\columnwidth}%
\begin{minipage}[c][1\totalheight]{0.02\columnwidth}%
\vspace{-0.5cm}
  (a)%
\end{minipage}%
\hspace{-0.1cm}%
\begin{minipage}[t][1\totalheight][c]{0.43\columnwidth}%
  \vspace{-0.4cm}
  \begin{center}\vspace{0cm}\includegraphics[scale=1.0]{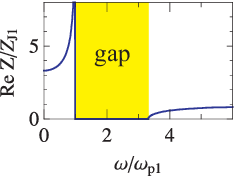}\par\end{center}%
\end{minipage}%
\hspace{0.6cm}%
\begin{minipage}[t][1\totalheight]{0.02\columnwidth}%
\vspace{-0.5cm}
  (b)%
\end{minipage}%
\hspace{-0.1cm}%
\begin{minipage}[t][1\totalheight][c]{0.41\columnwidth}%
  \vspace{-0.4cm}
  \begin{center}\vspace{0cm}\includegraphics[scale=1.0]{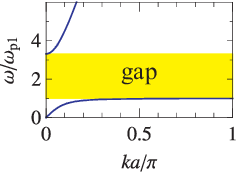}\par\end{center}%
\end{minipage}%
\end{minipage}%
\vspace{-0.2cm}
\par\end{centering}
\caption{ \label{fig:dispersionRelation} (Color online) (a) Real
part of the impedance (for an infinite array) and (b) dispersion
relation. Both show a gap in the same frequency range.  The
impedance is normalized by $Z_{\rm J1}=\sqrt{L_{\rm J1}/C_{\rm J1}}$
and the frequency by $\omega_{\rm p1}=1/\sqrt{L_{\rm J1}C_{\rm
J1}}$. The length of a unit cell is called $a$, and $k$ is the wave
number of a traveling wave solution. Here we used parameters $C_{\rm
J2}=C_{01}=0$, where the model reduces to that of one junction and
an additional inductance. Further, we chose parameters
$C_{02}/C_{\rm J1}=L_{\rm J2}/L_{\rm J1}=0.1$. }
\end{figure}

We briefly note another special case included in the two-junction 
model of Fig.~1c: If we choose $C_{01}=C_{\rm J2}=0$ and regard $L_{\rm J2}$ as
a geometric inductance instead of a Josephson inductance, we recover
a model containing one junction and an additional inductance in
series. This model was studied earlier, where the inductance $L_0$
was included in order to model the electromagnetic inductance of the
JJA transmission line\cite{jlt2000haviland}. This inductance led to
a gap in the real part of the impedance, taken between input port
and ground\cite{jlt2000haviland}, which corresponds to a gap in the
dispersion relation as shown in Fig.~\ref{fig:dispersionRelation}. However, for typical
parameters the gap appeared at approximately $10^{11}$-$10^{14}$ Hz.
While the lower frequency could, in principle, be reduced by
lowering the plasma frequency of the junction, the upper frequency
extended beyond the range of validity of the simple Josephson
junction model used. With the two-junction model presented here,
however, the upper band edge can be reduced in frequency by orders
of magnitude.

A geometric inductance $L_0\ll L_{\rm J}$ introduces a much stronger
asymmetry, which explains the wide gap and experimentally
inaccessibly high frequency of the upper band edge (which tends to
infinity for $L_0/L_{\rm J}\rightarrow 0$) in the model of
Ref.~\onlinecite{jlt2000haviland}. With the two-junction model, we
have a wide range of accessible parameters so that we can engineer
the band gap in an appropriate frequency range.

Our main results so far are Eqs. (1)-(5), which provide analytical solutions for
the dispersion relation and band edges for an infinite array with two-junction unit
cells in the linear regime, i.e., the 1d-JJA analog of a diatomic lattice. In the following we will consider
possibilities to achieve the required parameter space experimentally, and also discuss effects
of finite size, finite nonlinearity, and nonidentical junctions.

\section{Experimental considerations}
\subsection{Range of validity and experimental parameters}

The two branches in the dispersion relation and the associated gap
in frequency, where no propagating modes exist in the JJA
transmission line, could be a useful property for the design of
quantum circuits in the microwave region. The essential ingredient
for realizing this gap is an inequality of the parameters for each
of two junctions in the basis of the periodic structure, such that
Eq.~(\ref{eq:noGap}) is not fulfilled, or such that $C_{01}\neq C_{02}$. 
In this section we examine
realistic designs, subject to the constraints of fabrication, which
can achieve this asymmetry. The designs naturally fall into two
different parameter regimes depending on the transmission line
geometry used, coplanar or stripline.

When the JJA is made in a coplanar waveguide (CPW) geometry, where
the ground plane is on the sides of the JJA, the parameter regime
$C_{\rm J1},C_{\rm J2}\gg C_{01},C_{02}$  is easily realized.  In
this regime the gap vanishes when the plasma
frequencies of the two junctions become equal, and one should thus
aim for parameters $\omega_{\rm p1} \neq \omega_{\rm p2}$ in order
to have a gap. When fabricating JJAs, typically all junctions
are made in the same process step, resulting in a tunnel barrier
which is nearly uniform across the entire chip or wafer.  In this
case, the junction capacitance $C_{\rm J}$ and Josephson inductance $L_{\rm
J}$ will be proportional and inversely proportional to the junction
area, respectively, and the plasma frequency $\omega_{\rm
p}=1/\sqrt{L_{\rm J} C_{\rm J}}$ will therefore be independent of
the junction area. Thus, simply changing the junction area in the
fabrication process will not achieve $\omega_{\rm p1} \neq \omega_{\rm p2}$
which is required to have a gap.

Subject to the constraint of uniform tunnel barriers, there are two
ways to bring down the plasma frequency. The first method is to
increase $L_J$ of one of the junctions by forming a SQUID loop of
this junction and applying an external magnetic flux (see fig. 1a) . This method is
attractive because changing the external flux corresponds to tuning
the frequency range of the transmission gap. However, dropping the 
plasma frequency in this way also drops the critical
current of the transmission line, and therefore non-linear
corrections will become important at much lower power. The second
possibility to drop the plasma frequency is to fabricate an on-chip
capacitance in parallel with each junction.   This method will not
cause a degradation of critical current, however, it does require
more layers of lithography than the simple single layer process used
in the shadow deposition technique.  Fabrication with the Nb
trilayer technique however provides this parallel capacitance
naturally\cite{jap2005dolata}.

When designing an array in CPW geometry, one finds that the
characteristic impedance of the array is not well matched to the
termination impedance.  When the array is terminated with a direct
connection to an electrical lead, the termination of the array
impedance at microwave frequencies will be approximately $Z_0/2\pi =
60\Omega$, set by the free space impedance $Z_0=377$ $\Omega$. The
transmission line impedance $Z_{\rm A}$ of the JJA, which is the
pure real impedance of an infinite array, is in the zero frequency
limit given as $Z_{\rm A}(0)=\sqrt{(L_{\rm J1}+L_{\rm J2})/(C_{\rm
01}+C_{\rm 02})}$. It can be much larger than $Z_0$ in the CPW
geometry, where $C_0$ is relatively small, especially if $L_{\rm J}$
is made large by suppressing the critical current. We desire that
$Z_{\rm A} \ll R_{\rm Q} = h/4e^2 = 6.45$k$\Omega$, in order to
avoid quantum fluctuations of the phase which are not included in
our model based on classical phase dynamics. When $Z_{\rm A}\gg
R_{\rm Q}$, one finds that large quantum fluctuations of the phase
result in a Coulomb blockade, and our assumption of classical phase
dynamics has completely broken down\cite{prl1998chow}.

An alternative route to circuit design is based on the stripline
geometry, where the array is fabricated on top of a ground plane
with a thin, insulating (non-tunneling) barrier separating the array
islands from the ground plane.  For the stripline geometry, one
easily realizes the regime $C_{\rm J1},C_{\rm J2} \lesssim
C_{01},C_{02}$. In this regime, it is not necessary to have
different plasma frequencies of the two junctions, and we find that
a considerable gap in transmission also occurs if $C_{01}\neq
C_{02}$. In this case the gap appears well below the plasma
frequency. The condition $C_{01} \neq C_{02}$ is easily
realized in the stripline geometry when the junctions are made with 
overlapping films.

We have formulated a design for a JJA on an heavily oxidized Al
ground plane, to be fabricated with the shadow evaporation
technique.  In our Al tunnel junction fabrication, we find that it
is possible to achieve plasma frequencies as low as $\omega_{\rm
p1}/2\pi = \omega_{\rm p2}/2\pi = 33$ GHz.  A design with large area
base electrodes (2.5 $\mu$m x 20 $\mu$m) and long, narrow Dolan
bridges (0.1 $\mu$m x 2.5 $\mu$m) with small overlap (0.1 $\mu$m)
after shadow evaporation can achieve the following parameters:
$C_{01} = 0.68$~pF, $C_{02} = 3.4$~fF, $C_{J1} = 2.0$ pF, $C_{J2} =
12$ fF, with the array critical current being dominated by the
smaller junction 2, $I_{C2} = 170$ nA. For this design, we find that
the lower gap edge comes down in frequency to $\omega_{\rm
gL}/2\pi=8.5$ GHz, in a frequency range accessible to present day
qubit designs or broad band transmission measurements. For this
design, the transmission line impedance of the array is $Z_A(0)=53
\Omega$, which is well matched to the impedance of the input and
output ports of an array with high frequency leads connected at each
end. Such a design, with a rather low critical current and therefore
strongly nonlinear inductance, is ideal for the distributed
parametric amplifier\cite{apl1996yurke}. In the low power regime,
where linear behavior is expected, we find that such an array makes
a good superconducting low pass filter, with a very sharp drop in
transmission at 8.5 GHz in a design with only 20 unit cells in
series.

\subsection{Influence of parameter spread and transmission}
\label{subsec:parameterSpread}
\begin{figure}
\begin{centering}%
\hspace{-0.0cm}
  \includegraphics[scale=1.1]{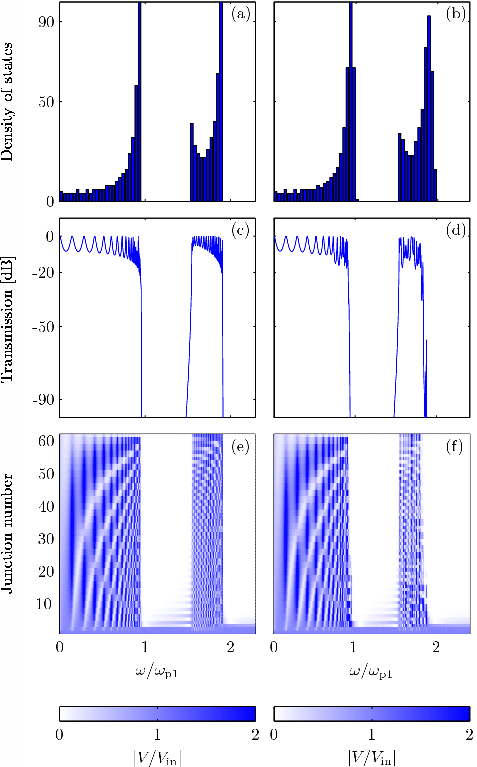}
\par\end{centering}
\caption{ (Color online) \label{fig:combined} (a),(b) Density of
states. (c),(d) Transmission. (e),(f) Voltage at the sites.  In
parts (b), (d), and (f), a $5\%$ standard deviation in Josephson
inductances is used. Further, $L_{\rm J2}=0.25 L_{\rm J1}$,
$C_{01}=C_{02}=0.2C_{1}=0.2C_{2}$, $Z_{\rm in}=Z_{\rm out}$, and
$Z_{\rm in}/Z_{\rm A}(0)=0.2$ .
 }
\end{figure}

Thus far we have assumed that the junctions can be fabricated identically. 
In reality there will be a spread of junction parameters in the 
fabrication, or disorder in the lattice. In one-dimension, even small disorder leads to Anderson localization\cite{pr1958anderson} of all states of an infinite system.  However, for weak disorder 
the localization length can be much larger than the finite-size array 
used in experiments.  Thus, we expect that the gap in the
spectrum, impedance, and transmission will persist provided that the
disorder is weak enough and the array is short enough that
localization effects can be ignored. We investigated these effects by 
numerical simulation, where the results are shown in 
Fig.~\ref{fig:combined}.  These results were obtained by classical circuit
theory, where we simulated a random spread of Josephson inductances
with normal distribution and standard deviation $5\%$ in
Fig.~\ref{fig:combined}b,d,f. This parameter spread, or disorder,
breaks the translational symmetry of the array, and wave vectors are no longer
well-defined.  However, it is still possible to investigate the
density of states, which is shown in Fig.~\ref{fig:combined}a,b.
Here we counted the number of states in a discrete frequency
interval, using periodic boundary conditions on an array with 500 unit cells, 
large enough to count a reasonable
number of states. Despite the spread in parameters, a gap can still
be clearly observed in the density of states. 
Localization effects cannot be observed from the density of states. However,
an investigation of the eigenmodes
shows localized states in the gap region, near the gap edge (not
shown).

In an experiment, it is easier to measure transmission than the
dispersion itself. Furthermore, accurate boundary conditions on a
finite length array become important for a real experiment. 
In Fig.~\ref{fig:combined}c-f we simulated an array with 30 unit cells and boundary conditions defined by the input and output leads
with transmission line impedances $Z_{\rm in}=Z_{\rm out}=50\Omega$.
Comparing Fig.~\ref{fig:combined}b and d, some states appearing at the upper band edges in the density of states are localized and do not contribute to the transmission.
The effect is however quite small for such short arrays.
We have also performed simulations with 500 junctions (not shown), where localized states appear near the band edge with higher probability.

The array simulated above behaves like a microwave resonator, even though it has 
direct electrical connection to the input and output terminals.  It
supports standing waves because the input and output impedances
$Z_{\rm in},Z_{\rm out}$ are not matched to the array impedance
$Z_A$. The standing waves can be seen by looking at the voltage at
each site as in Fig.~\ref{fig:combined}e,f. Note that because
$Z_A>Z_{\rm in}=Z_{\rm out}$, the voltage antinode occurs in the
middle of the array for the fundamental mode, opposite to standing
waves in resonators formed by a large point-like impedance at each
end of a transmission line, where $Z_{\rm in},Z_{\rm out}>Z_A$. Each
standing wave condition is associated with a peak in the
transmission as calculated in Fig.~\ref{fig:combined}c,d for the
case of no disorder, and 5\% parameter spread, respectively. For
frequencies inside the gap region, the transmission drops
drastically which can be understood by an exponential decay of
voltage amplitude from the edge of the array. However, the broad gap in transmission remains essentially
unaffected by a 5\% spread in parameters.  Thus, the design of such
a gap appears to be a robust and useful feature for quantum circuit
engineering.

\section{Summary and outlook}
We investigated the linear behavior of regular 1d-JJAs with
generalized unit cells, e.g., unit cells consisting of two junctions.  
The dispersion relation and the real part of the impedance show a gap in the 
dispersion relation, which is not present in arrays 
with only one junction per unit cell. We derived the
parameter dependence of the gap, and found that for a design with
two different Josephson junctions, the gap appears at frequencies of
the same order of magnitude as the plasma frequencies of the two
junctions. We suggested how to lower these frequencies in an
experimental setup in order to shift the gap to an accessible
frequency range, by replacing one of the two junctions per unit cell
with a SQUID, such that the gap can be tuned in situ, or by forming two different capacitances to ground.
The gap appears to be robust against a realistic parameter spread of the junctions (5\% standard deviation), and we have simulated a transmission experiment which we modeled with realistic
boundary conditions.

Our results could be used for comparatively simple
demonstration of tunable artificial crystals with Josephson
junctions. Such tunable artificial crystals could be used in 
circuit QED for frequency specific filters in qubit circuits. 
For example, by placing a  qubit in the middle of an array, 
when the qubit frequency lies inside the region
of the gap where no traveling modes are available, we expect the
relaxation of the qubit to be strongly suppressed. Decoherence of a
qubit is composed not only of the relaxation but also of the pure
dephasing, where the latter time scale is typically the critical,
shorter one. However, recent experiments reached extremely high
decoherence times, which, at least for part of the frequency range,
appeared to be limited by the relaxation\cite{prb2008schreier}. In
this case, suppression of the relaxation with a properly engineered
gap would allow refined studies on remaining sources of decoherence.

The model we presented in this paper is a linear analysis of the JJA
transmission line.  The interesting effects we describe arise due to
the plasma resonance of the Josephson junctions when they are
arranged in a discrete periodic structure. 
However, the linear approximation is valid only when the currents
flowing in the junctions are much less than the critical current.
When this condition is violated, nonlinear effects will appear,
which can be very strong in comparison with dissipative effects.  
These nonlinear effects give rise to a host of interesting
phenomena, such as parametric amplification\cite{apl2007tholen}.
Here, the ability to match the JJA transmission line impedance 
with the electromagnetic transmission line impedance, which is possible
for the case of stripline geometry, leads to the possibility of a broad
band parametric amplifier\cite{prl1988yurke}. Another interesting
nonlinear effect is the trapping or localization of energy in
discrete breather modes\cite{physToday2004campbell}, where a gap in
the dispersion relation is used to prevent the radiation damping of
Josephson oscillations in a junction in the middle of the
array\cite{commentBreathers}. We hope that the analysis presented
here will aid the development of future experiments in these
directions.\hspace{1cm}\\

\begin{acknowledgements}
We thank Hans Hansson for valuable discussions and comments on the
manuscript. This work was supported by the Swedish Research Council
(VR) and by NordForsk. DH gratefully acknowledges sabbatical support
from the Wenner-Gren Foundation.
\end{acknowledgements}

\appendix

\section{Derivation of the dispersion relation.}
The Lagrangian for the model consisting of unit cells with two types
of Josephson junctions (Fig.~\ref{fig:twoModels}b) is given as
\begin{eqnarray}
L_2&=&\sum_{j=0}^{N-1}[ \frac{C_{01}}{2}\dot{\Phi}_{j,1}^2
+\frac{C_{02}}{2} \dot{\Phi}_{j,2}^2]\nonumber\\
&&\hspace{-1cm}+\sum_{j=0}^{N-1}[ \frac{C_{\rm
J1}}{2}(\dot{\Phi}_{j-1,2}-\dot{\Phi}_{j,1})^2 +\frac{C_{\rm
J2}}{2}(\dot{\Phi}_{j,1}-\dot{\Phi}_{j,2})^2
]\nonumber\\
&&\hspace{-1cm}+\sum_{j=0}^{N-1}[E_{\rm
J1}\cos(\phi_{j-1,2}-\phi_{j,1}) +E_{\rm
J2}\cos(\phi_{j,1}-\phi_{j,2})]\ ,\nonumber\\
\end{eqnarray}
where we considered $N$ unit cells with periodic boundary
conditions, and where we eliminated the Josephson phases $\phi_{{\rm J},j}$ by generalized Kirchhoff constraints, and
defined fluxes and phases $\Phi_{j,1/2}=\Phi_0\phi_{j,1/2}/2\pi$ at
the capacitors to ground.

From this Lagrangian, one can find the equations of motion, and,
after the linear approximation, $\phi_{{\rm J},j}\ll 1$, make a
traveling wave ansatz,
\begin{equation}
\left(\begin{array}{c}
\Phi_{{\rm j},1}\\
\Phi_{{\rm j},2}\end{array}\right)=\left(\begin{array}{c}
u_k\\
v_k\, e^{ika/2}\end{array}\right)e^{i(k  ja-\omega t)}\ .
\end{equation}
Here, we introduced a length $a$ for the total unit cell, which
results in a factor $a/2$ for a single junction. The equations of
motion can be rewritten as a matrix $\mathbf F$ multiplying the vector
$(u_k,v_k)^T$ such that $\mathbf{F}(u_k,v_k)^T=0$. Nontrivial solutions
exist only when the determinant of $\mathbf{F}$ is zero, which results
in the dispersion relation stated in Eq.~(\ref{eq:dispTwo}). The
range of validity of the linear approximation is discussed 
in Appendix B.

\section{More general unit cells}
Here, we shall review the linear framework 
which we used to treat nonperiodic arrays 
as discussed in Section~\ref{subsec:parameterSpread}, and which can also be
used for more 
general types of unit cells, e.g. with three or more different
Josephson junctions or SQUIDs.
Finally, we discuss the effects of the nonlinear terms neglected so far.

{\bf Nonperiodic arrays. }
If we allow an arbitrary combination of capacitances, inductances
and Josephson junctions, the Lagrangian in the linear regime can
always be written as
\begin{equation}
\label{eq:lag_linear}
L=\sum_{\alpha,\beta=1}^M [\dot{\Phi}_\alpha
\frac{(\mathbf{C})_{\alpha\beta}}{2}\dot{\Phi}_\beta-\Phi_\alpha\frac{({\mathbf L^{-1}
})_{\alpha\beta}}{2}\Phi_\beta]\ .
\end{equation}
Here, $M$ is the total number of independent variables, and $\Phi_j$
are independent flux variables. In the case of the two-junction unit
cell we had $M=2N$, with $N$ the number of unit cells, and as
variables we used the integrated voltage at the capacitances to
ground, $\Phi_j=\int_{-\infty}^t V_j(t')dt'$. Further, $\mathbf C$
is the capacitance matrix and ${\mathbf L}^{-1}$ is the inverse
inductance matrix, which can contain both the kinetic inductance due
to Josephson junctions and the geometric inductance. This is a
problem of coupled harmonic oscillators, which can be diagonalized
by the transformation $\mathbf{\Phi}\rightarrow
\tilde{\mathbf{\Phi}}=\mathbf{U}^T\mathbf{C}^{1/2}\mathbf{\Phi}$.
Here we took into account that the matrices $\mathbf{C}$ and
$\mathbf{L}$ can always be chosen symmetric, and defined
$\mathbf{U}$ as the matrix which has columns consisting of the
normalized, real eigenvectors of the matrix
$\mathbf{\Omega}^2\equiv\mathbf{C}^{-1/2}\mathbf{L}^{-1}\mathbf{C}^{-1/2}$.
The transformed Lagrangian is given as $\label{eq:lagrangianDiag}
\tilde{L}=\frac{1}{2}\sum_{\lambda=1}^M [\dot{\tilde{\Phi}}_\lambda^2
-\omega_\lambda^2\tilde{\Phi}_\lambda^2]$, where we introduced the eigenvalues
$\omega_\lambda^2$ of $\mathbf{\Omega}^2$. These frequencies $\omega_\lambda$
resemble the dispersion relation for a regular array, and can still
be calculated in the presence of imperfections as used in section
III.B.

The equations of motion in this eigenbasis are decoupled and given
as $\ddot{\tilde{\Phi}}_\lambda=-\omega_\lambda^2 \tilde{\Phi}_\lambda$. The
Hamiltonian corresponding to the transformed Lagrangian is given as
$H=\frac{1}{2}\sum_\lambda(\tilde{Q}_\lambda^2+\omega_\lambda^2\tilde{\Phi}_\lambda^2)$,
where $\tilde{Q}_\lambda={\dot{\tilde{\Phi}}}_\lambda$ is the conjugate variable
to $\tilde{\Phi}$. For later convenience, this Hamiltonian can be
rewritten in the standard form $H=\sum_{\lambda=1}^N \hbar
\omega_\lambda(a_\lambda^\dagger a_\lambda+\frac{1}{2})$ when creation and
annihilation operators are defined by the equations
\begin{eqnarray}
\label{eq:PhiAndQ}
\tilde{\Phi}_{\lambda} & = &
\sqrt{\hbar/(2\omega_\lambda)}\left(a_{\lambda}^{\dagger}+a_{\lambda}\right)\
,\nonumber\\
\tilde{Q}_{\lambda} & = &
i\sqrt{\hbar\omega_\lambda/2}\left(a_{\lambda}^{\dagger}-a_{\lambda}\right)\ .
\end{eqnarray}

{\bf Periodic arrays  with extended unit cells. }
We consider now the condensed-matter like special case of Eq.~(\ref{eq:lag_linear}), where the $M$ degrees of freedom can be decomposed into a lattice with $N$ unit cells, each having a basis with 
$m$ degrees of freedom. The Lagrangian can then be rewritten as 
\begin{equation}
\label{eq:lag_linear2}
L=\sum_{j,l=1}^N\sum_{r,s=1}^m [\dot{\Phi}_{j,r}
\frac{(\mathbf{C})_{jl}^{(rs)}}{2}\dot{\Phi}_{l,s}-\Phi_{j,r}\frac{({\mathbf L^{-1}})_{jl}^{(rs)}}{2}\Phi_{l,s}]\ .
\end{equation}
The first index in $\Phi_{j,l}$ specifies the place in the lattice, and the second index specifies the
degree of freedom within the basis of this unit cell. Accordingly, we now have capacitance and inductance supermatrices, whose
lower indices act in the space of lattice places, and the upper indices act in the space of the basis.

This representation is 
useful in a periodic array with identical unit cells, such that
 $C_{jl}^{(rs)}=C_{j+x,l+x}^{(rs)}$ and ${(\mathbf L^{-1}})_{jl}^{(rs)}={(\mathbf L^{-1})}_{j+x,l+x}^{(rs)}$ for any $x\in \mathbb{Z}$. After the standard ansatz~\cite{book1976ashcroft} 
$(\Phi_{j,1},\Phi_{j,2},\dots,\Phi_{j,n})=\frac{1}{N}\sum_k e^{ika}(u_{k,1},u_{k,2},\dots,u_{k,n})$, the Lagrangian decouples due to the periodicity into
$L=\frac{1}{N}\sum_{k} L_k$
with
\begin{equation}
L_k= \sum_{r,s=1}^{m}\left[\dot{u}_{-k,r}\frac{C_{rs}(k)}{2}\dot{u}_{k,s}-u_{-k,r}\frac{{(\mathbf L^{-1})
}_{rs}(k)}{2}u_{k,s}\right].
 \end{equation}
For each wave vector $k$, a matrix structure as
in Eq.~(\ref{eq:lag_linear}) remains. The band structure for given $k$ can then be obtained in analogy to Appendix A or by explicit diagonalization,   
where the relevant matrices are now defined as $C_{r,s}(k)\equiv \sum_{l=1}^{N} e^{ikla} C_{1,1+l}^{(r,s)} $ and ${(\mathbf L^{-1})}_{r,s}(k)
\equiv \sum_{l=1}^{N} e^{ikla} {(\mathbf L^{-1})}_{1,1+l}^{(r,s)} $.
Diagonalization is achieved by a transformation as above, $\tilde{\mathbf{u}}\mathbf{=\mathbf{U^{\dagger}\ C}^{1/2}\ u}$, and creation and annihilation operators are introduced
by $\tilde{u}_{k,n}=\sqrt{\hbar/(2\omega_{k,n})}(a_{-k,n}^{\dagger}+a_{k,n})
$.

{\bf Nonlinearities. }
In our derivation of the dispersion relation, we approximated terms
of form $\cos(\phi_\alpha-\phi_\beta)$ by expanding to the quadratic term in the phases
or fluxes (linear in the equation of motion). Here we shall briefly consider the
fourth order terms, which more generally can be of form $L_{\rm nl}=\sum_{\alpha}\frac{E_{\rm J\alpha}(2\pi/\Phi_0)^4}{4!}(\sum_{\beta} \gamma_{\beta}^{(\alpha)}\Phi_\beta)^4$, where
the matrix elements $\gamma_\beta^{(\alpha)}$ take values $1$, $-1$, and $0$ only. We calculate the leading order correction to the dispersion due to these terms for a weak nonlinearity.
After transformation to the eigenbasis of the linearized system
and making use of Eq.~(\ref{eq:PhiAndQ}),
the nonlinearity can be brought to the form
\begin{equation}
\label{eq:fourthOrder}
H_{\rm nl}=-\sum_{\lambda,\mu,\nu,\sigma}f_{\lambda\mu\nu\sigma}(a_{\lambda}+a_{\lambda}^{\dagger})(a_{\mu}+a_{\mu}^{\dagger})(a_{\nu}+a_{\nu}^{\dagger})(a_{\sigma}+a_{\sigma}^{\dagger}),
\end{equation}
with coefficients $f_{\lambda\mu\nu\sigma}$ to be discussed below.

We consider a weak nonlinearity and a monotonic drive with frequency $\omega_{\rm d}$ such that
the system response is dominated by the linear behaviour, and predominantly one mode $k$
in a specific band $n$ is highly excited. 
Due to the huge population $N_\lambda$ of this state $\lambda=(k,n)$,  most interactions will be between this state and at most one other state.
Taking
into account energy conservation in the sense of a rotating wave approximation, we can approximate the nonlinear part of the Hamiltonian as
\begin{eqnarray}
H_{\rm nl}&=&-6f_{\lambda\lambda\lambda\lambda}a_\lambda^\dagger a_\lambda^\dagger a_\lambda a_\lambda-24\sum_{\mu, \mu\neq \lambda}
f_{\mu\mu\lambda\lambda}a_\lambda^\dagger a_\lambda a_\mu^\dagger a_\mu\nonumber\ .\\
\end{eqnarray}
The Heisenberg equation of motion then reads
$i\hbar\dot{a}_\lambda=
(\hbar\omega_\lambda-12f_{\lambda\lambda\lambda\lambda} a_\lambda^\dagger a_\lambda -24\sum_{\mu, \mu\neq \lambda}
f_{\mu\mu\lambda\lambda}  a_\mu^\dagger a_\mu) a_\lambda\ .
$

Note that other states than $\lambda$ can in general be excited due to the nonlinearity, which mixes different modes. However, 
if the population $N_\lambda$ of mode $\lambda$ is much higher than the population of all other modes together, only the first term above is relevant, and leads within
a semiclassical approximation to a frequency shift in the dispersion relation $\omega(\lambda)\rightarrow \omega(\lambda)-12f_{\lambda\lambda\lambda\lambda}N_\lambda$, provided that 
$f_{\lambda\lambda\lambda\lambda}N_{\lambda}\ll \hbar \omega_\lambda$. To fulfill both the latter condition and the assumptions of the semiclassical approximation, we have to stay in the regime
$f_{\lambda\lambda\lambda\lambda}\ll \hbar\omega_\lambda/N_{\lambda} \ll \hbar\omega_\lambda$. 
While one can change $N_\lambda$ via the drive strength, the quantity $f_{\lambda\lambda\lambda\lambda}$ is given
from the device geometry, and derived from the capacitive energies and the (potentially tunable) Josephson energies.

In general an expression for $f_{\lambda\lambda\lambda\lambda}$ becomes complicated and is best obtained numerically, but for a periodic array with a one-junction unit cell 
(and thus only one band, $n=0$)
one can obtain a simple
analytical expression. 
Taking into account that the frequency of the linear term is given as $\hbar\omega_k=\sqrt{8E_{\rm C_0}E_{\rm J}\beta_k/(1+\beta_k C_{\rm J}/C_0)}$, where $\beta_k=4\sin^2(ka/2)$
and $E_{\rm C_0}=e^2/2C_0$, we find 
$f_{k,-k,k,-k}=-(\hbar\omega_k)^2/24NE_{\rm J}$.
Note that taking into account the $E_{\rm J}$ dependence of the frequency, the strong Josephson energy cancels out, and the nonlinearity is in this regime dependent on the charging energy scales,
i.e., on the distribution of the capacitances.
The condition of a weak nonlinearity above ($f\ll \hbar\omega$) thus requires that $\hbar\omega_k\ll E_{\rm J}$, which for said $\omega_k$ is equivalent to $E_{C},E_{C_0}\ll E_{\rm J}$. This requirement can also be expressed in engineering terms as $Z_A\ll R_Q$, as stated in the main text. 
While we only calculated the prefactor of the nonlinearity $f_{k,-k,k,-k}$ explicitly for an array with a single junction per unit cell, we do not expect conceptual changes for the validity of the linear
regime when two junctions per unit cell are present.

\end{document}